**Communication**

# Stable covalently photo-cross-linked poly(ionic liquid) membrane with gradient pore size

Alessandro Dani,[1] Karoline Täuber,[1] Weiyi Zhang,[1] Helmut Schlaad,[2]* Jiayin Yuan[1,3]*

\_\_\_\_\_\_\_\_\_\_

[1]Dr. A. Dani, Dr. K. Täuber, W. Zhang, Prof. J. Yuan
Department of Colloid Chemistry, Max Planck Institute of Colloids and Interfaces
Am Mühlenberg 1, 14476 Potsdam, Germany
[2]Prof. H. Schlaad
Institute of Chemistry, University of Potsdam
Karl-Liebknecht-Str. 24-25, 14476 Potsdam, Germany
E-mail: schlaad@uni-potsdam.de
[1,3]Prof. J. Yuan
Department of Chemistry & Biomolecular Science and Center for Advanced Materials Processing
Clarkson University, 8 Clarkson Avenue, Potsdam 13699, USA
E-mail: jyuan@clarkson.edu

\_\_\_\_\_\_\_\_\_\_

An imidazolium-based poly(ionic liquid) is covalently cross-linked *via* UV light-induced thiol-ene (click) chemistry to yield a stable porous polyelectrolyte membrane with gradients of cross-link density and pore size distribution along its cross-section.

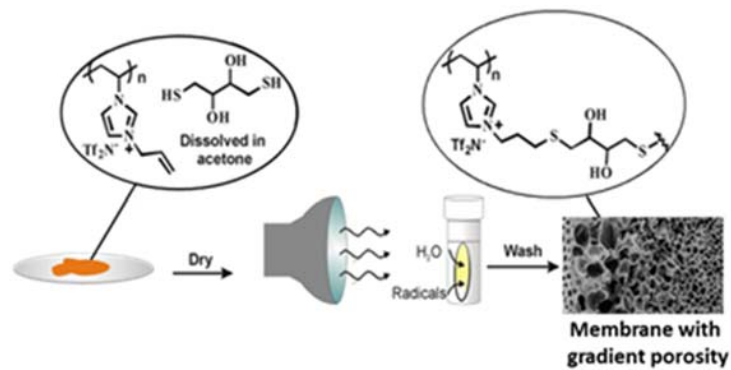

# 1. Introduction

Polyelectrolyte membranes have become an emerging class of materials in the past years.[1, 2] They can be made up from one or two different polyelectrolytes, whose charged nature allows for a mutual inter-polyelectrolyte complexation that leads to membrane formation.[3-5] By exploiting different kinds of polyelectrolytes, it is possible to tune the membrane properties for various applications, such as separation, sensing, or catalysis.[6-13] Moreover, by use of templating agents, layer-by-layer deposition, or phase separation, membranes can be developed for pores of different sizes and distributions.[1, 6] A promising type of polyelectrolyte membrane is the one based on poly(ionic liquid)s (PILs), in which the ionic-liquid based moieties along polymer chains can functionalize the membrane and open up routes to a large variety of applications, such as separations, actuation, pH sensing, and many others. [14-24]

The current method used to create porosity in PIL membranes exploits PIL phase separation from a non-solvent during inter-polyelectrolyte complexation. Since the complexation is triggered by a base (usually ammonia) that diffuses across the membrane in an asymmetrical way, it is possible to achieve a gradient in cross-linking degree and pore size.[18, 19, 25-28] Due to the dynamic nature of the ionic bonding in the PIL film, this pore production methodology produces porous membranes that are instable under the harsh conditions encountered when they are used as corrosion barriers, separation membranes, medical implants, or components in electrochemical devices.[29-31] In fact, when exposed to high ionic strength solutions, these PIL membranes typically undergo partial or full morphological degradation of the porous structure, mainly due to the dissociation of the electrostatic cross-links. In order to obtain stable, porous polyelectrolyte membranes, covalent cross-links between polyelectrolyte chains would be preferred over electrostatic cross-links. Typically, polyelectrolyte membranes are covalently cross-linked by means of a post-polymerization treatment that involves photoreactions, condensation reactions, or click-chemistry reactions in order to introduce additional covalent cross-links in addition to the electrostatic ones.[32-36]

In this communication, we describe a direct way to obtain a porous polyelectrolyte membrane by covalent cross-linking of a single PIL polyelectrolyte, avoiding the use of inter-polyelectrolyte complexation. The covalent cross-links among the PIL chains were synthesized by a thiol-ene photo-induced reaction in a non-solvent to the PIL, [37-39] resulting in phase separation that induces porosity in the as-formed membrane. Furthermore, this cross-linking method leads to a pore size gradient along the membrane cross section, producing a stable, asymmetric, covalently cross-linked PIL membrane in a single step procedure.

## 2. Results and Discussion

This fabrication procedure for making porous PIL membranes *via* thiol-ene cross-linking is illustrated in Scheme 1. The single PIL component, poly(3-allylmethyl-1-vinylimidazolium bromide) was synthesized *via* quaternization of poly(*N*-vinyl imidazole) with 1-allyl bromide according to our previous work.[40] Next, the bromide anion was exchanged to bis(trifluoromethane sulfonyl)imide anion ($Tf_2N^-$) to obtain the hydrophobic PIL, poly(3-allylmethyl-1-vinylimidazolium $Tf_2N$) (PAMVIm-$Tf_2N$) that will be phase-separated from water. More synthetic experimental details are given in the Supporting Information. To fabricate the porous membrane, PAMVIm-$Tf_2N$ (103.8 mg, 0.25 mmol of the repeat units) and dithioerythritol (DTE; 38.6 mg, 0.25 mmol) were dissolved in acetone (1 mL), and stirred until a homogeneous solution formed. This solution was then cast onto a glass plate (2 cm$^2$), and the solvent was evaporated at room temperature overnight. The dry film was then immersed into an aqueous solution of a radical photo-initiator (Irgacure® 2959, 25 mg in 5 mL). The reactor was immediately placed in an ice bath and irradiated by an UV-A light source (225 mW/cm$^2$) for 2 h in ambient atmosphere. Note, the surfaces of the film in contact with the aqueous solution and the glass plate are named as TOP and BOTTOM surfaces, respectively. A free-standing porous membrane was then peeled off from the glass plate and rinsed with water to remove any unreacted DTE and initiator. During this process, the initiator radicals (and water molecules as well) diffuse from the aqueous solution into the PIL/DTE blend film, triggering the thiol-ene click-reaction to cross-link the PIL chains. Hydrophobic PAMVIm-Tf$_2$N is insoluble in water; hence, the simultaneous penetration of water into the polymer film induced phase separation, resulting in the formation of pores. Therefore, the cross-linking reaction and phase separation process run parallel to produce and stabilize the freshly formed porous structure. The obtained membrane was insoluble in acetone, a good solvent for its precursor PAMVIm-Tf$_2$N, confirming successful covalent cross-linking. In a control experiment in the absence of the Irgacure® 2959 photo-initiator to see whether or not UV light alone would be sufficient to cross-link the PIL,[32] no porous membrane was formed and the treated polymer/DTE blend film remains soluble in acetone.

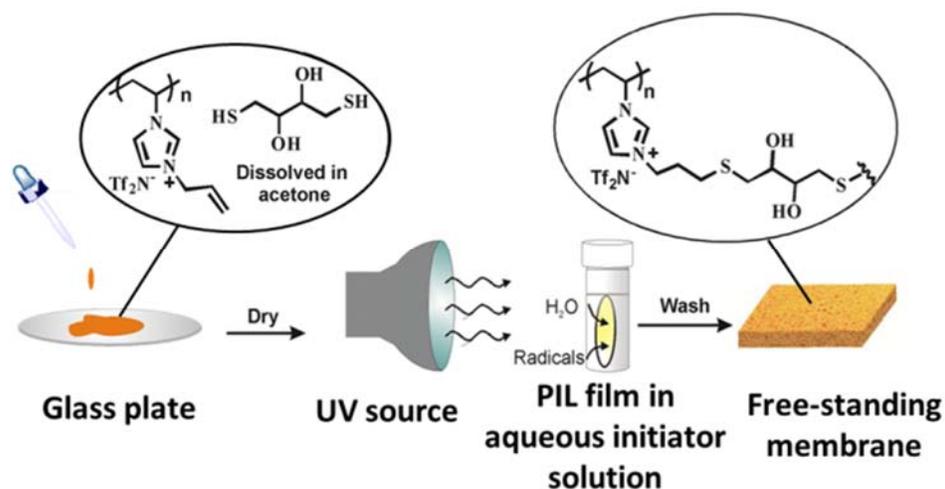

**Scheme 1.** Illustration of the membrane fabrication process. A solution of PAMVIm-Tf$_2$N and DTE in acetone was dropped onto a glass plate, after which upon solvent evaporation a dry PIL/DTE blend film was formed and immersed in an aqueous solution of a radical photo-initiator. Irradiation with a UV light source induces cross-linking and pore formation, yielding a free-standing porous membrane.

The porous structure of the thiol-ene cross-linked membrane was investigated by scanning electron microscopy (SEM). As shown in Figure 1a, micrometer-sized pores are formed that are densely packed, creating a fully interconnected pore structure across the entire membrane. A porosity gradient, which arises from the asymmetric penetration of the photo-initiator across the membrane, is well recognizable in the membrane cross-section. The top of the membrane exhibits smaller pores because the pores are stabilized at an early stage of the phase separation by the fast cross-linking reaction at high concentration of radicals. The bottom part of the membrane, on the other hand, contains less radicals–mainly due to a diffusion effect– to stabilize the pore at a later stage of the phase separation, ultimately leading to bigger pores. The average pore size, as calculated by statistical analysis of the SEM image considering all pores along the gradient, is (3 ± 1) µm (see Figure 1b). In contrast, when the PIL film was thermally cross-linked without DTE in the film (see the synthesis section in the Supporting Information), the obtained membrane contained interconnected pores with a much larger size of (8 ± 4) µm (Figure S1). This difference in pore sizes arises from the fact that the kinetics of thermal cross-linking is much slower than phase separation. In addition, the photochemically cross-linked membrane could be used to separate polystyrene latex particles of 100 nm in size from micron-sized particles (Figure S2), thus demonstrating its applicability as separation membrane.

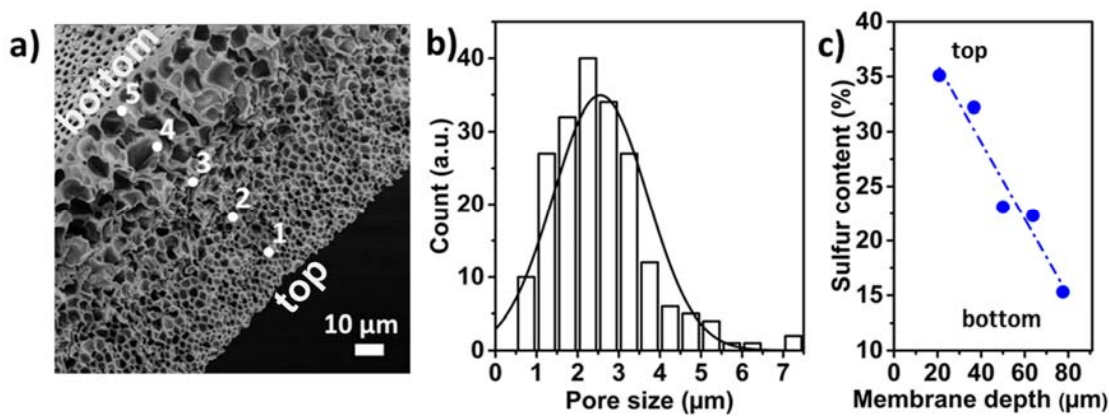

**Figure 1.** (a) SEM image of the porous cross-section of the membrane and (b) pore size distribution obtained from statistical analysis on 200 measured pores. (c) EDX analysis profile of the sulfur content along the membrane depth from the top.

Energy-dispersive X-ray (EDX) analysis performed along the membrane cross-section showed that the content of sulfur decreases progressively from 35.1% on the top to 15.3% at the bottom of the membrane (see Figure 1c and Table S1 in the Supporting Information). These values can be attributed to a decreasing amount of DTE from top to bottom of membrane cross section, which is related to a decreasing degree of cross-linking over the membrane cross-section. It is important to note that the theoretical sulfur content in DTE and PAMVIm-Tf$_2$N are 41.6 wt% and 15.4 wt%, respectively. EDX analysis shows a little overestimation of sulfur content at a high weight ratio, possibly due to a concurrent self-coupling reaction of DTE in presence of air. The gradient in cross-linking degree enables the membrane to be responsive to organic media. For example, a wet membrane in water will undergo actuation in contact with acetone due to the different swelling degree of the membrane from top to bottom (see Video S1 in the Supporting Information).

ATR-FT-IR spectra of the membrane were collected before and after the thiol-ene reaction on both sides of the membrane. The spectrum collected on the membrane before the thiol-ene reaction, top red curve in Figure 2, is a superposition of the vibrational bands of PAMVIm-TF$_2$N and of DTE; both reference spectra are shown in Figures S2-S3 (in the Supporting Information). The band related to the C-S stretching of the formed thioether is found in the 705-570 cm$^{-1}$ range. However since this band is usually very weak in FT-IR spectroscopy, it is not reliable for monitoring the thiol-ene reaction. Instead, the S-H stretching band of DTE at 2554 cm$^{-1}$ was monitored, because it is visible on the membrane before UV treatment and vanished completely after the reaction.[41] Interestingly, the C-H stretching regions are different on opposite sides of the membrane. The FT-IR spectrum from the top side, faced to the UV light source (Figure 2, middle green curve), shows the asymmetric and symmetric C-H stretching bands at 2935 cm$^{-1}$ and 2872 cm$^{-1}$, respectively, which are ascribed to the thioether methylene group.[42] On the bottom side of the membrane (Figure S3, bottom blue curve), these two bands appear weaker, indicating fewer thioether bonds and thus a lower degree of cross-linking. Hence, ATR-FT-IR analysis qualitatively confirms the successful thiol-ene cross-linking as well as different cross-linking densities along the membrane cross section. All other vibrational features in the low-frequency region, which are assigned to the polymer backbone and the $Tf_2N^-$ anion, were unaltered after the thiol-ene reaction.

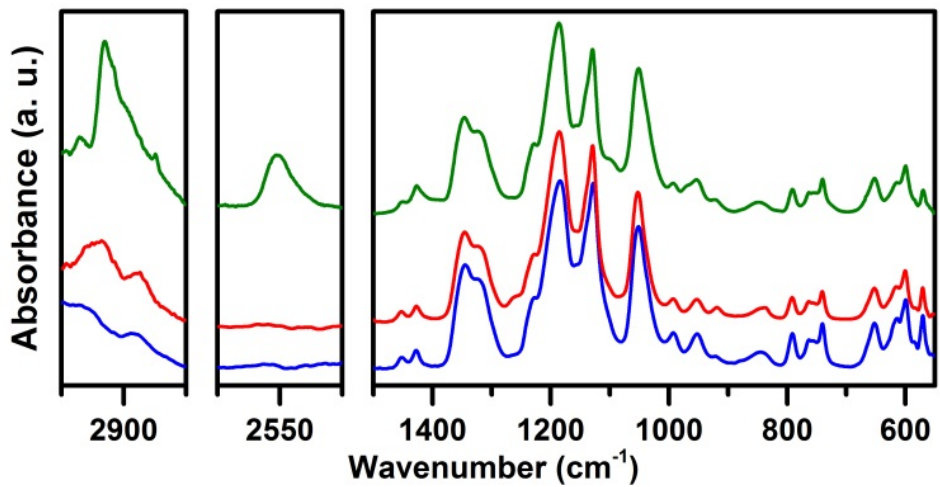

**Figure 2.** ATR-FT-IR spectra of the membrane before (green curve) and after the thiol-ene reaction on the top surface (red curve) and the bottom surface (blue curve).

In order to evaluate the thermal stability of the thiol-ene cross-linked membrane, thermogravimetric analysis was performed on the membrane and the two starting components, PAMVIm-$Tf_2N$ and DTE (see Figure S5 in Supporting Information). The pristine PIL is the most thermally stable component, showing a 2 wt% mass loss at 323 °C and 10 wt% at 371 °C when analyzed under dry $N_2$ using a heating ramp of 10 °C/min. On the contrary, DTE is rather thermally unstable under these testing conditions, showing a 2 wt% mass loss at 124 °C and 10 wt% at 175 °C. The PIL membrane obtained from PAMVIm-$Tf_2N$/DTE shows a small decomposition step at around 180 °C under these conditions, probably due to the decomposition of some residual unreacted DTE molecules in the membrane. For the PIL membrane sample, 2 wt% mass loss was observed at 179 °C, and 10 wt% mass loss is at 367 °C. Its decomposition profile resembles the pure PIL sample. Additionally, the ATR-FT-IR spectrum virtually did not change after annealing the membrane sample at 280 °C (see Figure S6 in Supporting Information), suggesting that the small weight loss has no impact on the general chemical structure of the membrane. It should be noted that the membrane in a wet state is flexible (see Figure S7 in Supporting Information).

## 3. Conclusions

In conclusion, this work describes a simple way to fabricate a covalently cross-linked, porous PIL membrane having gradient cross-link density and pore size distribution along its cross section. Notably, such a membrane structure has not yet been reported for covalently cross-linked polyelectrolyte membranes. The gradient in cross-linking density can be used for an asymmetrical swelling when the membrane is exposed to solvent, resulting in a bending movement. The membrane chemical structure is based on covalent cross-links and not electrostatic cross-links, paving the way to create polyelectrolyte membranes bearing freely adjustable anions, aimed to expanded applications, such as separator in batteries, separations and filtration in different chemical environment.

## Supporting Information

Supporting Information is available from the Wiley Online Library or from the authors

Acknowledgements: The authors acknowledge financial support from the Max Planck Society, and European Research Council (ERC) starting grant NAPOLI-639720.

Received: Month XX, XXXX; Revised: Month XX, XXXX; Published online:

Keywords: poly(ionic liquid), membrane, photo-cross-linked, porous structure

# Supporting Information

DOI: 10.1002/marc.2013#####

[1] Klitzing, R. V.; Tieke, B. Polyelectrolyte Membranes. *Adv. Polym. Sci.* **2004**, *165*, 177.
[2] D. L. Gin, R. D. Noble, *Science* **2011**, *332*, 674.
[3] N. Joseph, P. Ahmadiannamini, R. Hoogenboom, I. F. J. Vankelecom, *Polym. Chem.* **2014**, *5*, 1817.
[4] E. V. Konishcheva, U. E. Zhumaev, W. P. Meier, *Macromolecules* **2017**, *50*, 1512.
[5] V. Mikhalevich, C. Zelmer, A. Car, C. Palivan, W. Meier, "Bio-inspired Polymer Membranes", in *Bio-inspired Polymers*, The Royal Society of Chemistry, 2017, p. 221.
[6] Q. Zhao, Q. F. An, Y. Ji, J. Qian, C. Gao, *J. Membr. Sci.* **2011**, *379*, 19.
[7] L. Ouyang, D. M. Dotzauer, S. R. Hogg, J. Macanás, J.-F. Lahitte, M. L. Bruening, *Catal. Today* **2010**, *156*, 100.
[8] H. Gu, F. Yan, J. Texter, *Macromol. Rapid Commun.* **2016**, *37*, 1218.
[9] J. Grünauer, S. Shishatskiy, C. Abetz, V. Abetz, V. Filiz, *J. Membr. Sci.* **2015**, *494*, 224.
[10] F. Gu, H. Dong, Y. Li, Z. Sun, F. Yan, *Macromolecules* **2014**, *47*, 6740.
[11] I. Tokarev, M. Orlov, S. Minko, *Adv. Mater.* **2006**, *18*, 2458.
[12] J. E. Bara, C. J. Gabriel, E. S. Hatakeyama, T. K. Carlisle, S. Lessmann, R. D. Noble, D. L. Gin, *J. Membr. Sci.* **2008**, *321*, 3.
[13] N. W. DeLuca, Y. A. Elabd, *J. Polym. Sci., Part B: Polym. Phys.* **2006**, *44*, 2201.
[14] B. A. Voss, J. E. Bara, D. L. Gin, R. D. Noble, *Chem. Mater.* **2009**, *21*, 3027.
[15] J. E. Bara, S. Lessmann, C. J. Gabriel, E. S. Hatakeyama, R. D. Noble, D. L. Gin, *Ind. Eng. Chem. Res.* **2007**, *46*, 5397.
[16] X. Fan, H. Liu, Y. Gao, Z. Zou, V. S. J. Craig, G. Zhang, G. Liu, *Adv. Mater.* **2016**, *28*, 4156.
[17] Y. Kohno, D. L. Gin, R. D. Noble, H. Ohno, *Chem. Commun.* **2016**, *52*, 7497.
[18] Q. Zhao, M. Yin, A. P. Zhang, S. Prescher, M. Antonietti, J. Yuan, *J. Am. Chem. Soc.* **2013**, *135*, 5549.
[19] Q. Zhao, J. W. C. Dunlop, X. Qiu, F. Huang, Z. Zhang, J. Heyda, J. Dzubiella, M. Antonietti, J. Yuan, *Nat. Commun.* **2014**, *5*, 4293.
[20] L. C. Tomé, A. S. L. Gouveia, C. S. R. Freire, D. Mecerreyes, I. M. Marrucho, *J. Membr. Sci.* **2015**, *486*, 40.
[21] L. C. Tome, D. Mecerreyes, C. S. R. Freire, L. P. N. Rebelo, I. M. Marrucho, *J.f Mater. Chem. A* **2014**, *2*, 5631.
[22] L. C. Tome, I. M. Marrucho, *Chem. Soc. Rev.* **2016**, *45*, 2785.
[23] L. Ansaloni, J. R. Nykaza, Y. Ye, Y. A. Elabd, M. Giacinti Baschetti, *J. Membr. Sci.* **2015**, *487*, 199.
[24] J. Lu, F. Yan, J. Texter, *Prog. Polym. Sci.* **2009**, *34*, 431.
[25] Q. Zhao, J. Heyda, J. Dzubiella, K. Täuber, J. W. C. Dunlop, J. Yuan, *Adv. Mater.* **2015**, *27*, 2913.
[26] K. Täuber, Q. Zhao, M. Antonietti, J. Yuan, *ACS Macro Lett.* **2015**, *4*, 39.
[27] K. Täuber, A. Zimathies, J. Yuan, *Macromol. Rapid Commun.* **2015**, *36*, 2176.
[28] Thünemann, A. F.; Müller, M.; Dautzenberg, H.; Joanny, J.-F.; Löwen, H. Polyelectrolyte Complexes. *Adv. Polym. Sci.* **2004**, *166*, 113.
[29] D. V. Andreeva, E. V. Skorb, D. G. Shchukin, *ACS Appl. Mater. Interfaces* **2010**, *2*, 1954.
[30] Z. Tang, Y. Wang, P. Podsiadlo, N. A. Kotov, *Adv. Mater.* **2006**, *18*, 3203.
[31] X. Li, S. De Feyter, D. Chen, S. Aldea, P. Vandezande, F. Du Prez, I. F. J. Vankelecom, *Chem. Mater.* **2008**, *20*, 3876.
[32] P. Ott, K. Trenkenschuh, J. Gensel, A. Fery, A. Laschewsky, *Langmuir* **2010**, *26*, 18182.
[33] L. Shan, H. Guo, Z. Qin, N. Wang, S. Ji, G. Zhang, Z. Zhang, *RSC Adv.* **2015**, *5*, 11515.
[34] G. K. Such, J. F. Quinn, A. Quinn, E. Tjipto, F. Caruso, *J. Am. Chem. Soc.* **2006**, *128*, 9318.

[35] J. J. Harris, P. M. DeRose, M. L. Bruening, *J. Am. Chem. Soc.* **1999**, *121*, 1978.
[36] I. Pastoriza-Santos, B. Schöler, F. Caruso, *Adv. Funct. Mater.* **2001**, *11*, 122.
[37] C. E. Hoyle, T. Y. Lee, T. Roper, *J. Polym. Sci., Part A: Polym. Chem.* **2004**, *42*, 5301.
[38] C. E. Hoyle, C. N. Bowman, *Angew. Chem. Int. Ed.* **2010**, *49*, 1540.
[39] N. K. Singha, H. Schlaad, "Thiol–ene Based Functionalization of Polymers", in *Functional Polymers by Post-Polymerization Modification*, Wiley-VCH Verlag GmbH & Co. KGaA, 2012, p. 65.
[40] J. Yuan, A. G. Marquez, J. Reinacher, C. Giordano, J. Janek, M. Antonietti, *Polym. Chem.* **2011**, *2*, 1654.
[41] N. Sheppard, *Trans. Faraday Soc.* **1950**, *46*, 429.
[42] N. B. Colthup, L. H. Daly, S. E. Wiberley, "Chapter 12 - Compounds containing boron, silicon, phosphorus, sulfur, or halogen", in *Introduction to Infrared and Raman Spectroscopy (Third Edition)*, Academic Press, San Diego, 1990, p. 355.

# Stable covalently photo-cross-linked poly(ionic liquid) membrane with gradient pore size

Alessandro Dani, Karoline Täuber, Weiyi Zhang, Helmut Schlaad,* Jiayin Yuan*

__________

Dr. A. Dani, Dr. K. Täuber, W. Zhang, Prof. J. Yuan
Department of Colloid Chemistry, Max Planck Institute of Colloids and Interfaces
Am Mühlenberg 1, 14476 Potsdam, Germany
Prof. H. Schlaad
Institute of Chemistry, University of Potsdam
Karl-Liebknecht-Str. 24-25, 14476 Potsdam, Germany
E-mail: schlaad@uni-potsdam.de
Prof. J. Yuan
Department of Chemistry & Biomolecular Science and Center for Advanced Materials Processing
Clarkson University, 8 Clarkson Avenue, Potsdam 13699, USA
E-mail: jyuan@clarkson.edu

__________

**Materials and characterization methods**
All reagents were purchased from Sigma-Aldrich and used without any further purification. Solvents were of analytical grade.
$^1$H NMR and $^{13}$C{$^1$H} NMR spectra were measured on a Bruker DPX-400 spectrometer operating at 400.1 MHz ($^1$H) and 100.6 MHz ($^{13}$C); DMSO-$d_6$ was used as the solvent. ATR-FTIR measurements were performed with Nicolet iS5 FT-IR spectrometer from ThermoFisher Scientific on a single reflection ATR diamond. The measurement range was from 4000 cm$^{-1}$ to 500 cm$^{-1}$ (in steps of 2 cm$^{-1}$); 128 scans were collected for the background and 64 scans for sample spectra. Size exclusion chromatography (SEC) with simultaneous UV and RI detection was performed in DMSO at 70 °C using a column set of two 300 x 8 mm$^2$ PSS-GRAM columns with porosities of 10$^2$ and 10$^3$ Å; calibration was done with pullulan standards. Thermogravimetric measurements were performed on a TG 209 F1 Libra from Netzsch. Samples were analyzed from room temperature to 600 °C at a heating rate of 2 K·min$^{-1}$ under a nitrogen flow of 30 mL·min$^{-1}$ and using a platinum crucible. Scanning electron microscopy (SEM) was performed on a JEOL JSM7500F microscope at 10 kV equipped with EDX detector from Oxford Instrument. Samples were coated with gold before examination.

**Synthesis of poly(3-allylmethyl-1-vinylimidazolium bromide)**
Poly(3-allylmethyl-1-vinylimidazolium bromide) (PAMVIm-Br) was synthesized according to previous work (Yuan *et al.*, Polym. Chem., 2011, **2**, 1654-1657). The quaternization of polyvinylimidazole ($M_w^{app}$ 180 Da, dispersity 1.9, by SEC) was confirmed by $^1$H NMR by the integration of peaks at 6.16, 5.41 and 4.81 ppm related to the allyl group; quaternization yield: 98%. The calculated $M_w^{app}$ for PAMVIm-Br was 410 kDa. $^1$H NMR (DMSO-$d_6$) δ(ppm) = 9.73 (1H), 8.40-7.68 (2H), 6.19 (1H), 5.43 (2H), 4.83 (3H).

**Synthesis of poly[3-allylmethyl-1-vinylimidazolium bis(trifluoromethane sulfonyl)imide]**
Poly(3-allylmethyl-1-vinylimidazolium bromide) (2 g; 9.29 mmol of repeating units) was dissolved in water (200 mL). Lithium bis(trifluoromethane sulfonyl)imide (4 g; 13.95 mmol) was dissolved in water (10 mL) and added dropwise to the polymer solution. The precipitated poly[3-allylmethyl-1-vinylimidazolium bis(trifluoromethane sulfonyl)imide] (PAMVIm-Tf$_2$N) was filtered off and washed extensively with water. The polymer was then freeze-dried and isolated in form of a white powder. $^1$H NMR (DMSO-$d_6$) δ(ppm) = 8.92 (1H), 8.68-7.74 (2H), 5.86 (1H), 5.51 (2H), 4.63 (3H).

**Synthesis of photo-crosslinked PAMVIm-Tf$_2$N membrane**
PAMVIm-Tf$_2$N (103.8 mg, 0.25 mmol of repeating units) and dithioerythritol (DTE) (38.6 mg, 0.25 mmol) were dissolved in acetone (1 ml), and the homogenous solution was cast onto a glass plate (2 cm$^2$). Acetone was evaporated at room temperature, and the dry film was immersed into a solution of water (5 mL) containing the photoinitiator 2-hydroxy-4′-(2-hydroxyethoxy)-2-methylpropiophenone (Irgacure® 2959) (25 mg). The reactor was immediately placed in an ice bath at 8 cm in front of a gas discharge lamp (ECE Series UV lamp from DYMAX with bulbs emitting primarily in UV-A region, 225 mW/cm$^2$) and irradiated for 2 hours. The product was isolated from the glass substrate and washed with water to remove residual DTE and initiator.

**Synthesis of thermally crosslinked PAMVIm-Tf$_2$N membrane**
PAMVIm-Tf$_2$N (103.8 mg, 0.25 mmol of repeating units) was dissolved in acetone (1 ml), and the homogenous solution was cast onto a glass plate (2 cm$^2$). Acetone was evaporated at room temperature, and the dry film was immersed into a solution of water (5 mL) containing **2,2'-azobis[2-methyl-*N*-(2-hydroxyethyl)propionamide] (**VA-086) (50 mg) as a thermal initiator. The mixture was heated at 90 °C in nitrogen atmosphere for 16 hours. The product was isolated from the glass substrate and washed with water to remove residual initiator.

**Filtration experiments**
The membrane was attached to a hollow glass cylinder with a hole in the bottom (Ø 1.5 mm). The glass cylinder was then put into an empty syringe and solvents could pass through the device. The glass cylinder was added in order to make sure that the solvent does not flow through the syringe wall without passing by the membrane. In order to keep the weight of the solution on top of the membrane constant (thus constant pressure to the membrane), a separation funnel was placed above the syringe setup and the solution flow into the syringe was adjusted to keep a constant weight.

**Table S1.** EDX analyses on C, S, F, O and N performed on spots along membrane cross-section.

| spot | % C | % S | % F | % O | % N |
|------|------|------|-----|------|-----|
| 1 | 40.5 | 35.1 | 7.5 | 9.0 | 7.9 |
| 2 | 43.7 | 32.2 | 8.0 | 10.2 | 6.9 |
| 3 | 48.7 | 23.1 | 9.8 | 10.6 | 8.9 |
| 4 | 50.4 | 22.3 | 9.9 | 10.0 | 8.8 |
| 5 | 54.6 | 15.3 | 9.7 | 10.5 | 9.9 |

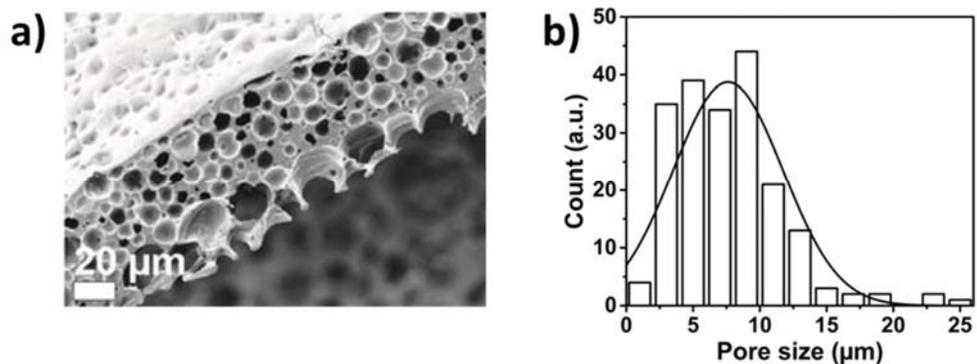

**Figure S1.** (a) SEM image of the porous cross-section of membrane obtained by thermal radical crosslinking of PAMVIm-Tf$_2$N (in the absence of DTE), and (b) pore size distribution obtained from the statistical analysis on 200 measured pores.

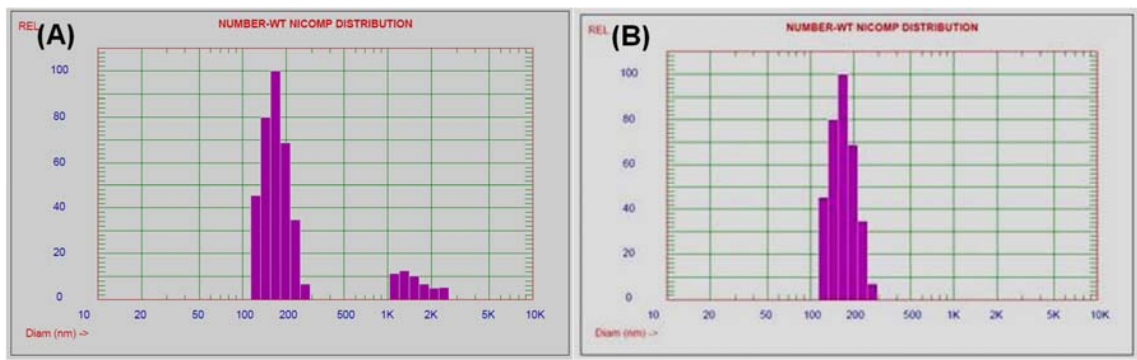

**Figure S2.** Number-averaged particle size distributions of PS particles before (A) and after (B) the filtration through the as-synthesized membrane.

- 14 -

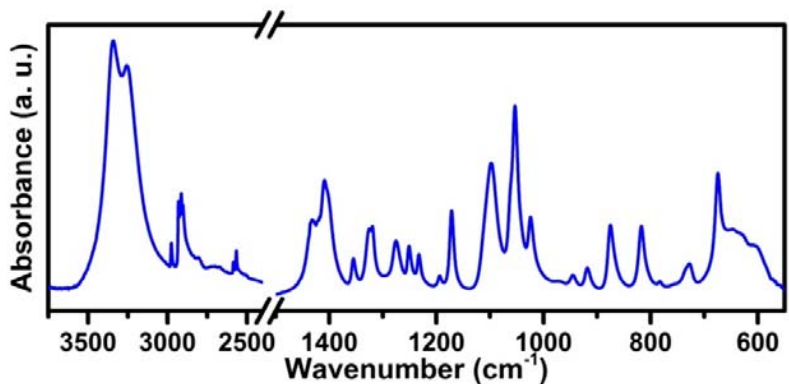

**Figure S3.** ATR-FTIR spectrum of dithiothreitol (DTE).

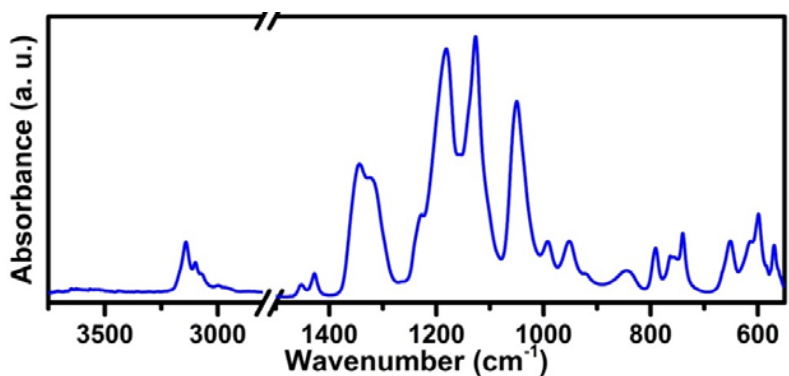

**Figure S4.** ATR-FTIR spectrum of poly(3-allylmethyl-1-vinylimidazolium Tf$_2$N).

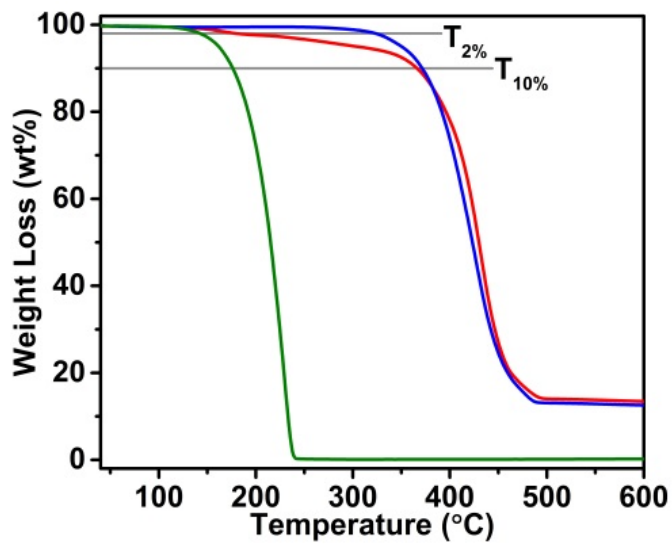

**Figure S5.** Thermogravimetric profiles of DTE (green curve), PAMVIm-Tf$_2$N (blue curve) and the thiol-ene cross-linked membrane (red curve).

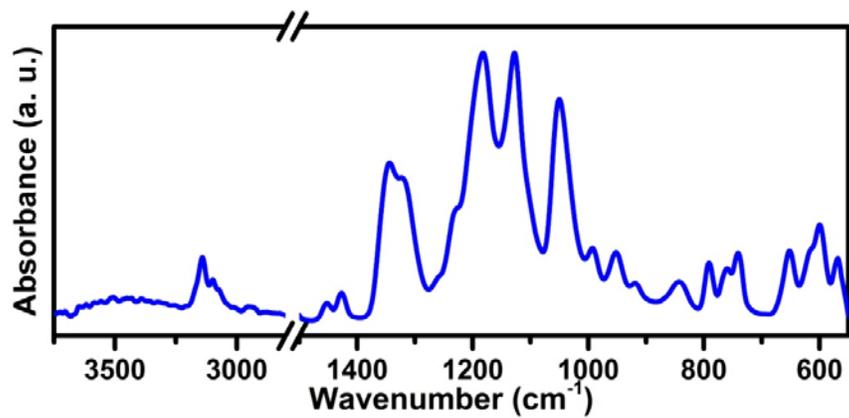

**Figure S6.** ATR-FTIR spectrum of thiol-ene crosslinked PIL membrane treated at 280 °C under nitrogen.

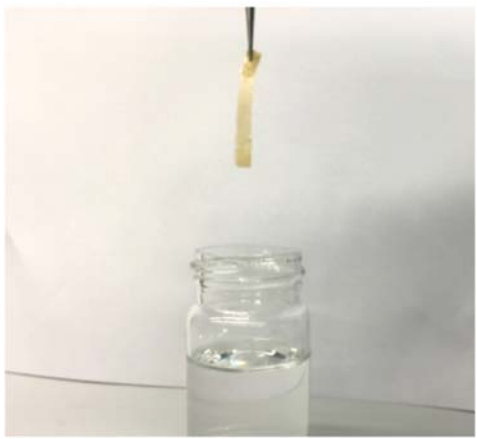

**Figure S7.** Photograph of a rectangular membrane stripe.